\documentclass[sigconf]{acmart}
\usepackage[utf8]{inputenc}

\AtBeginDocument{%
  \providecommand\BibTeX{{%
    \normalfont B\kern-0.5em{\scshape i\kern-0.25em b}\kern-0.8em\TeX}}}

\usepackage{multirow}
\usepackage{listings}
\usepackage[backgroundcolor=blue!10,bordercolor=gray]{todonotes}

\begin{document}

\title[Exploring the Security Awareness of the Python and JavaScript Open Source Communities]{Exploring the Security Awareness of the\\Python and JavaScript Open Source Communities}

\author{G\'abor Antal}
\affiliation{%
  \institution{University of Szeged}  
  \city{Szeged}
  \country{Hungary}}
\email{antal@inf.u-szeged.hu}

\author{M\'arton Keleti}
\affiliation{%
  \institution{University of Szeged}  
  \city{Szeged}
  \country{Hungary}}
\email{keletim@inf.u-szeged.hu}

\author{P\'eter Heged\H{u}s}
\affiliation{%
  \institution{MTA-SZTE Research Group on Artificial Intelligence}  
  \city{Szeged}
  \country{Hungary}}
\email{hpeter@inf.u-szeged.hu}


\begin{abstract}
Software security is undoubtedly a major concern in today's software engineering.
Although the level of awareness of security issues is often high, practical experiences show that neither preventive actions nor reactions to possible issues are always addressed properly in reality.
By analyzing large quantities of commits in the open-source communities, we can categorize the vulnerabilities mitigated by the developers and study their distribution, resolution time, etc. to learn and improve security management processes and practices.
  
With the help of the Software Heritage Graph Dataset, we investigated the commits of two of the most popular script languages -- Python and JavaScript -- projects collected from public repositories and identified those that mitigate a certain vulnerability in the code (i.e. vulnerability resolution commits).
On the one hand, we identified the types of vulnerabilities (in terms of CWE groups) referred to in commit messages and compared their numbers within the two communities.
On the other hand, we examined the average time elapsing between the publish date of a vulnerability and the first reference to it in a commit.
  
We found that there is a large intersection in the vulnerability types mitigated by the two communities, but most prevalent vulnerabilities are specific to language.
Moreover, neither the JavaScript nor the Python community reacts very fast to appearing security vulnerabilities in general with only a couple of exceptions for certain CWE groups.
\vspace{-5pt}
\end{abstract}




\maketitle

\vspace{-5pt}
\section{Introduction}\label{sec:introduction}

Software security is one of the most striking problems of today's software systems.
Large impact security vulnerabilities are explored on a daily basis, for example, a serious flaw~\cite{sudo-macos} has been discovered in 'Sudo', a powerful utility used in macOS this February.
Security problems can cause not just financial damage~\cite{Anderson2013}, but can compromise vital infrastructure, or used to threaten entire countries.

Our focus in this paper is to examine vulnerability mitigation (i.e. corrective code changes to resolve security vulnerabilities) within the open-source community and their typical types.
We specifically target Python and JavaScript open-source projects as these languages are very popular and widely used in many domains today.
By getting a detailed picture of what security vulnerabilities and when are mitigated in the open-source community of these languages, we can identify vulnerability categories that are not sufficiently addressed, explore patterns that might help to build more efficient vulnerability prediction models, or even discover some patterns that may help in generalizing the models.
We investigate the following two research questions in this work:

\emph{RQ1: What are the typical security vulnerability types the JavaScript and Python open-source communities mitigate and how do they relate to each other?}

\emph{RQ2: How quickly the JavaScript and Python open-source communities mitigate a newly published security vulnerability?}

Based on the rich set of data in the Software Heritage Graph Dataset~\cite{MSR20DC}, we found that the JavaScript projects refer to security vulnerabilities falling into 87 different categories, the Python projects to 71, out of which 55 security vulnerability categories are common.
For vulnerability categorization, we use the widely adopted Common Weakness Enumeration (CWE) list~\cite{cwe}.
Despite the large intersection in the security vulnerability types, the number of mitigated vulnerabilities differ significantly depending on the language of the projects.
For example, Cross-site Scripting (CWE-79), Path Traversal (CWE-22), Improper Input Validation (CWE-20) and Uncontrolled Resource Consumption (CWE-400) type of vulnerabilities are mitigated mostly in JavaScript projects, while Resource Management Errors (CWE-399) and Permissions, Privileges, and Access Controls (CWE-264) are mitigated mostly in Python.

The growing number of vulnerability mitigating commits is a common tendency in both languages, but it is proportionate to the growth of the total number of commits.
The vulnerability mitigation per total commit ratio increases only slowly, however, there was a significant increase in the amount of vulnerability mitigation in the year 2018 for both JavaScript and Python projects (see Figure~\ref{fig:count_year_distr}).
Regarding the number of days elapsing between the publish date of a particular security vulnerability and the date of the first commit with its mitigation is varying to a large extent.
Typically, Python commits mitigate vulnerabilities no older than 100 days, while some JavaScript commits mitigate vulnerabilities older than a year.

\vspace{-10pt}
\section{Approach}\label{sec:approach}

Our approach is based on collecting the vulnerability mitigation commits for JavaScript and Python projects from the dataset, which are potentially connected to a public CVE~\cite{cve} entry.
To achieve this, we used a very simple but effective heuristics-based approach, similar to those widely used in works related to bug data collection~\cite{GVS19, 10.1145/1083142.1083147}.
First, we searched for the commits containing the patterns ``CVE-'', ``CWE-'', ``NVD-'' (all of them are case insensitive) in their commit messages using SQL queries.
Referring to a CVE or CWE identifier in the commit message is a widely used practice in case of vulnerability mitigation patches, so the community can understand why the given commit is extremely important and urgent to be merged.
By filtering the \texttt{revision} table, we created a temporal table called \texttt{cve\_revs} with 357,757 rows (from the original 1.26 billion rows).

After the first filtering step, we had to identify the programming language of the project a given commit belongs to.
Since the structure of the database did not provide an effective way to do this, we used the information retrieved from the revisions' root directory:

\begin{itemize}
    \item We considered a revision as a Python revision if its root directory contained either \texttt{\_\_init.py\_\_} or \texttt{setup.py}.
    Without at least one of these files, the project cannot be used as a Python module~\cite{packagingpython, pilgrim2009dive, younker2009foundations} (nor published on PyPI~\cite{pypi}), therefore it is a viable heuristics to detect Python projects.
    \item We considered a revision to be a JavaScript one if its root directory contained either \texttt{index.js}, \texttt{app.js}, \texttt{server.js} as one of these files will most likely be included in the root directory~\cite{nodejsdocumentation} of a JavaScript project.
    We did not consider \texttt{package.json} for identifying a revision as a JavaScript revision because \texttt{package.json} is often used in other languages as well, such as PHP (e.g. Symfony uses \texttt{package.json} to manage tools that are necessary for packing the application's frontend~\cite{symfonydocs}).
\end{itemize}

Based on this second round of filtering, we got 3,718 rows for Python and 4,136 rows for JavaScript, which we stored in two new tables: \texttt{cve\_revs\_py} and \texttt{cve\_revs\_js}, respectively.
We analyzed the data collected in these instead of the original \texttt{revision} table.


\subsection{Tools and Queries for Data Mining}


We processed the collected Python and JavaScript revisions using Python scripts and pandas~\cite{mckinney2011pandas},
and used regular expressions\footnote{$(CVE-\backslash d\{4\}-\backslash d\{4,\})$, $(CWE-[\backslash d]\{1,4\})$, and $(NVD~.+)$} to find and extract the CVE/CWE IDs from the commit messages.
All the used regular expressions and extraction scripts for finding CVE/CWE and vulnerability mitigating revisions are available in our online asset package.\footnote{\url{https://doi.org/10.5281/zenodo.3699486}}
We also tried to filter commits for ``NVD'', but there were no matching commit messages.
If a commit message contained more than one CVE or CWE references, we extracted all and considered them separately (i.e. the commit contained mitigation for more than one vulnerabilities).
As a commit message can contain the same CVE/CWE IDs several times (for example, it can be in the first line of the commit message and later it can appear in the description as well), we had to remove the duplicates.
Thus one CVE/CWE entry is considered only once per revision.

Several rows has not been filtered out in the first step, but in the processing step we could not find any CVE/CWE IDs in their commit messages.
We examined and validated all of these cases by hand.
These revisions contained messages that could pass our first filtering but did not mention any valid CVE/CWE IDs, for example, \textit{execve-safe, Glennvd-patch-1, nvd-downloader, no CVE-id}.

As we focused on the types of vulnerabilities, which can be described by the CWE identifier of the security problem category the vulnerability belongs to, we had to link each CVE entries to the corresponding CWE categories of the vulnerabilities.
To achieve this, we relied on the data provided by the National Vulnerability Database~\cite{nvd} and used a customized version of CVE manager by Atlasis~\cite{cve-manager} to parse the JSON data files describing the CVE entries with meta-information, like its corresponding CWE category.
Besides the CWE group of a CVE entry, we also extracted the publishing date, severity, and the base impact score of every CVE entries.

Some revisions contained references to CWE groups without mentioning any specific CVE entries.
These revisions were mapped directly to the referenced CWE categories.

\vspace{-5pt}
\subsection{Software Heritage Graph Dataset Version}
We performed our study using the compressed PostgreSQL format~\cite{MSR19SH} of the full Software Heritage Graph Dataset~\cite{MSR20DC}.
It took us several tries to correctly import the dataset into a local database.
With some modifications to the original load script (e.g. removing concurrent index creation), we managed to import the whole database into a local server.

The technical specifications of the database server we used were 20-core Intel CPU (2,6 GHz), 90 Gbs of RAM, 5 Tb SSD.
Despite the quite strong hardware, the data import and queries were rather slow due to the enormous size of the database.
To speed up the process, we created intermediate tables from the relevant information in a filtered and transformed way.

\vspace{-5pt}
\section{Results}\label{sec:results}

\begin{figure}[htb!]
\vspace{-8pt}
\centering
\includegraphics[width=0.7\columnwidth]{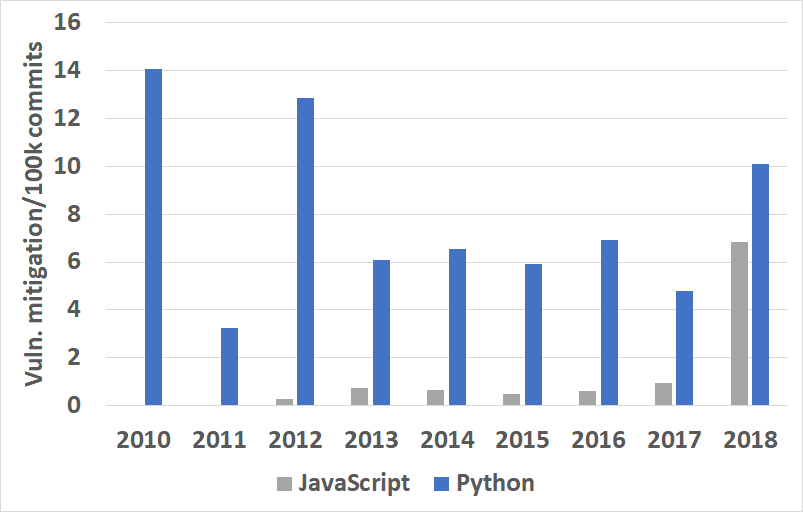}
\vspace{-10pt}
\caption{Vulnerability mitigation ratio per year}
\label{fig:count_year_distr}
\vspace{-10pt}
\end{figure}

After all filtering steps, we identified a total number of 3,458 vulnerability mitigation commits (i.e. commit messages containing valid CVE or CWE IDs) for JavaScript and 2,884 for Python to which we were able to resolve the corresponding CWE security type groups as well.
Figure~\ref{fig:count_year_distr} shows the ratio of commits over the years in terms of the average number of vulnerability mitigation commits per 100k commits.

\begin{figure*}
\centering
\includegraphics[width=1.52\columnwidth]{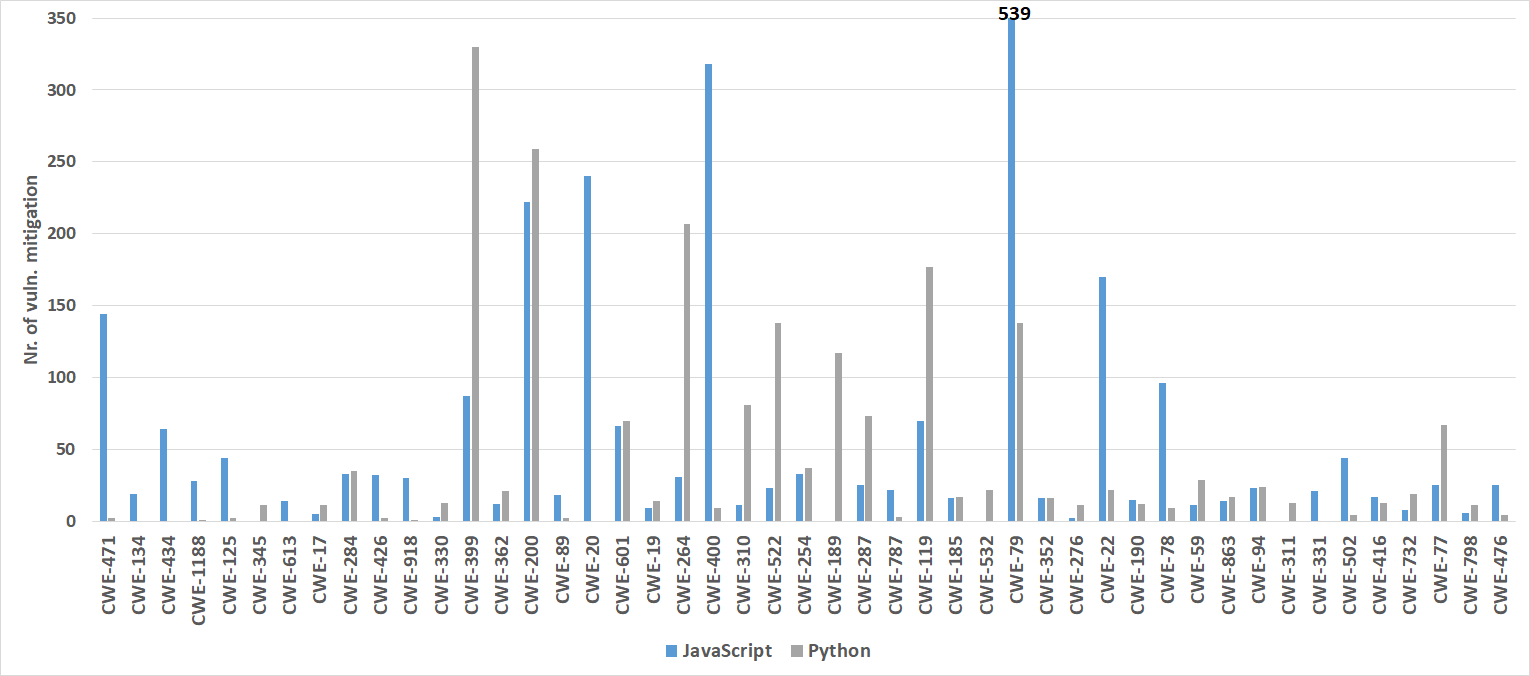}
\vspace{-10pt}
\caption{Number of security issues found with the given CWE types}
\label{fig:cwe_counts}
\vspace{-15pt}
\end{figure*}

While Python vulnerability mitigation ratio is quite stable, the same ratio for JavaScript projects grows consistently from 2015, with a large peak in 2018, but is still lower than that of Python projects.
As there are no JavaScript commits in Software Heritage Dataset before 2010, and the data for 2019 is still incomplete, we omitted those years from the analysis.
Table~\ref{tab:stat} provides further details on the number of detected vulnerability mitigation commits and the total number of commits in the analyzed years.
The distribution of the referenced CWE vulnerability types are depicted in Figure~\ref{fig:cwe_counts}.

\begin{table}[htbp]
  \vspace{-12pt}
	\centering
	\footnotesize
  \caption{Commit statistics per year}
	\vspace{-10pt}
    \begin{tabular}{r|r|r|r|r}
    \hline
    \multicolumn{1}{l|}{Year} & \multicolumn{1}{c|}{Vuln. JS} & \multicolumn{1}{c|}{Vuln. PY} & \multicolumn{1}{c|}{Total JS} & \multicolumn{1}{c}{Total PY} \\
    \hline
    \hline
    2010  & 0     & 225   & 102,525 & 1,597,160  \\
    2011  & 0     & 67    & 675,492 & 2,068,155  \\
    2012  & 6     & 343   & 2,078,887 & 2,663,836  \\
    2013  & 41    & 209   & 5,705,696 & 3,436,804  \\
    2014  & 84    & 291   & 12,692,836 & 4,440,660  \\
    2015  & 111   & 328   & 23,794,463 & 5,537,294  \\
    2016  & 239   & 453   & 38,990,699 & 6,527,350  \\
    2017  & 393   & 329   & 40,883,417 & 6,835,803  \\
    2018  & 2584  & 639   & 37,729,971 & 6,315,866  \\
		\hline
    \end{tabular}%
  \label{tab:stat}%
	\vspace{-19pt}
\end{table}%

\subsection{Typical Security Issue Types (RQ1)}

To answer RQ1, we examined the extracted vulnerability mitigation commits with 103 different CWE categories.
From these 103, 55 CWE types occurred in both JavaScript and Python commit messages, while 32 CWE groups were found only in JavaScript projects, while 16 only in Python projects (however, the number of vulnerabilities with such types were very low).

We examined the most popular CWE categories in more detail.
The CWEs having at least 150 references in either of the analyzed languages are as follows:
\begin{itemize}
    \item \emph{CWE-79} -- Improper Neutralization of Input During Web Page Generation (Cross-site Scripting).
    \item \emph{CWE-399} -- Resource Management Errors.
    \item \emph{CWE-200} -- Information Exposure.
    \item \emph{CWE-20} -- Improper Input Validation.
    \item \emph{CWE-264} -- Permissions, Privileges, and Access Controls.
    \item \emph{CWE-400} -- Uncontrolled Resource Consumption.
    \item \emph{CWE-119} -- Improper Restriction of Operations within the Bounds of a Memory Buffer.
    \item \emph{CWE-22} -- Improper Limitation of a Path-name to a Restricted Directory (Path Traversal).
\end{itemize}

Interestingly, except for CWE-200 that is the type of the vulnerabilities mitigated in more than 200 commits in both languages, each of the other six CWE groups can be attributed to either JavaScript or Python projects (i.e. one of the languages contain the majority of the mitigation to these vulnerability types).
On the one hand, Cross-site Scripting (CWE-79), Path Traversal (CWE-22), Improper Input Validation (CWE-20) and Uncontrolled Resource Consumption (CWE-400) type of vulnerabilities are mitigated mostly in JavaScript projects.
All these vulnerability types are primarily relevant for web applications, where JavaScript is heavily used at the client-side, thus it is more probable that a JavaScript project encounters such vulnerabilities.
On the other hand, mitigation of Resource Management Errors (CWE-399), Permissions, Privileges, and Access Controls (CWE-264), and Improper Restriction of Operations within the Bounds of a Memory Buffer (CWE-119) type of vulnerabilities occur in Python commits mostly.
These are more relevant at the server-side, where Python seems to dominate.
There is a significant overlap in these categories as well, so projects from both languages have vulnerabilities with all these CWE types, but based on the data we have, it seems that these are more typical for a particular language.

\vspace{-8pt}
\subsection{Reaction Times to Security Issues (RQ2)}

\begin{figure}[htb!]
\centering
\vspace{-7pt}
\includegraphics[width=0.8\columnwidth]{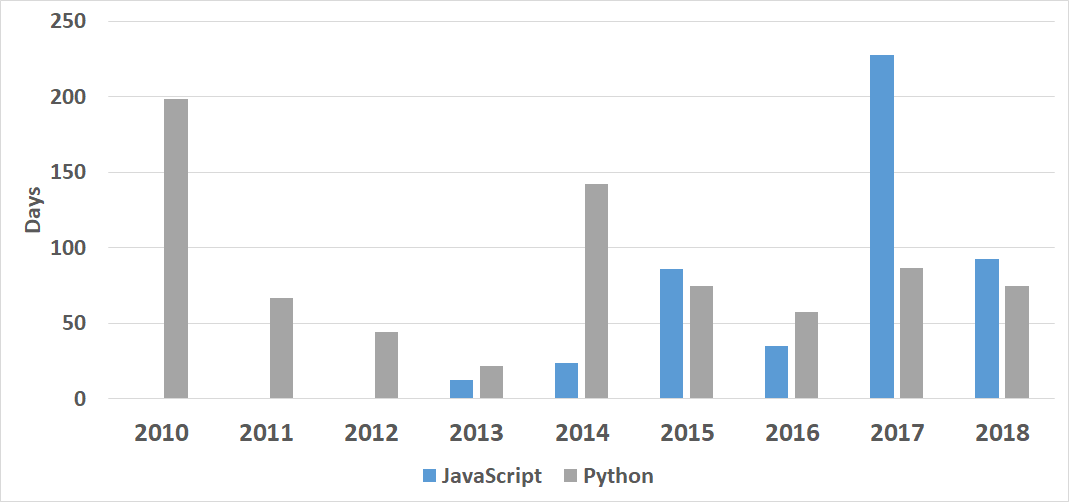}
\vspace{-10pt}
\caption{Average number of days between mitigation commit date and CVE publish date grouped by years}
\vspace{-5pt}
\label{fig:avg_fix_days_y}
\end{figure}

To answer RQ2, we analyzed the average number of days elapsing between a mitigation commit date and the publish date of a CVE entry mentioned in that commit.
Figure~\ref{fig:avg_fix_days_y} depicts a general overview of these average number of days per year.
We can see that it takes about 100 days on average for both communities to start mitigating a public vulnerability in their code-base, with some peaks in years 2010 and 2014 for Python and 2017 for JavaScript.
Therefore, we can conclude that at a very general level, neither the JavaScript nor the Python communities react fast to appearing vulnerabilities in their code.
It would be also interesting to see, if there are reported CVE entries that are never mitigated in reality, but it would require an entirely different methodology and could be a good future research.

\begin{figure}[htb!]
\vspace{-7pt}
\centering
\includegraphics[width=0.7\columnwidth]{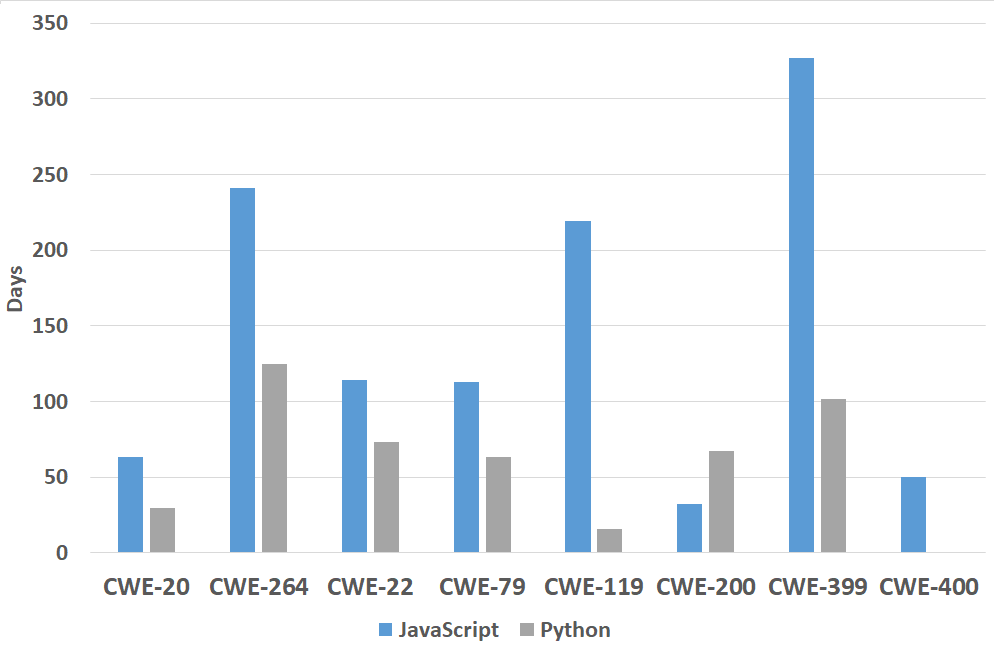}
\vspace{-10pt}
\caption{Average number of days between commit date and issue publish date for the most common CWEs}
\label{fig:fix_days_top_cwe}
\vspace{-8pt}
\end{figure}

We also examined the eight most prevalent CWE categories from the same aspect.
The average number of days elapsed between the publish date of a CVE entry and the date of its mitigation commit for the top eight CWEs are shown in Figure~\ref{fig:fix_days_top_cwe}.
The Python community reacts 1.5-14 times faster to these type of vulnerabilities than the JavaScript community; most of the mitigation commits appear 50 days or less after the publish date of the corresponding vulnerabilities.
In the case of JavaScript, only vulnerabilities from three CWE categories enjoy extra care (CWE-20, CWE-200, CWE-400), all the others are mitigated after at least 100 days.

The JavaScript community reacts exceptionally fast to information exposure (CWE-200) type of vulnerabilities (after only 32.5 days on average), while improper input validation (CWE-20) and uncontrolled resource consumption (CWE-400) are mitigated after about 50 days on average.
Interestingly, the vulnerabilities falling into the CWE categories characteristic to Python (CWE-264, CWE-399, and CWE-119) are mitigated after 200 days or more.

\section{Related Work}\label{sec:related}

By exploring the life-cycle of the vulnerabilities~\cite{10.1145/1162666.1162671, 6227141}, one can understand their nature better, which helps to find or predicting them.
Analogously to general bug prediction models, specific vulnerability prediction models have been introduced~\cite{10.1145/3196398.3196454, 10.1145/1853919.1853925}.
A big question regarding them is how well they generalize across projects (or even across languages)~\cite{10.1145/2832987.2833051}.

In their work, Li et Paxson~\cite{frankli2017} conducted a large-scale empirical study (with over 4000 security patches) to investigate the vulnerability fix  development life cycle and its characteristics, compared to the non-security bug fixing life cycle and characteristics.
They revealed that third of all security fixes are introduced more than 3 years after publishing.

Xu et al.~\cite{xu2017spain} proposed a binary-level patch analysis framework called SPAIN, which identifies security (and non-security) patches by analyzing the binaries.
The framework also detects patch and vulnerability patterns that can be used to detect similar patches/vulnerabilities in the given binaries.
In contrast to this work, we analyzed the source code changes mitigating vulnerabilities.

V{\'a}squez et al.~\cite{Vsquez2017AnES} analyzed 660 Android-related vulnerabilities and their fixes.
They used both NVD and Google Android security bulletins to collect their data.
Their analysis include vulnerability types and the hierarchical relationship between vulnerabilities, the impacted components and the survivability of the vulnerabilities.
We instead analyzed JavaScript and Python vulnerabilities.


\section{Threats to Validity}\label{sec:threats}

We had to apply heuristics to determine the language of the projects (as the exact solution would have been practically infeasible due to the database structure).
Due to this, we might have omitted some projects as well as identified some projects wrongly.
However, as our heuristics are based on widely established guidelines and best practices that most of the projects follow, the number of these projects should be minimal.

In most of the cases the committers mention CVE IDs explicitly, however, there are unusual references, for example ``Fixed XSS (with CVE number 2020-100)'' or ``CVE-2020-20500/330/34/345''.
Also, there is a chance that a committer mentions a CVE in a context that is not related to fixing its underlying security issue.
In such cases, we might drop valid vulnerability mitigation commits or include invalid ones.
To estimate the impact of this threat, we manually evaluated 700 randomly selected commit messages from the identified revisions. 
In the vast majority of the evaluated cases, the commit messages refer to CVE IDs as we anticipated, thus the impact of this threat should be minimal.


\vspace{-5pt}
\section{Conclusions}\label{sec:conclusions}

Using the Software Heritage Graph Dataset, we analyzed the vulnerability mitigation commits in the Python and JavaScript projects from two aspects.
On the one hand, we identified the types of vulnerabilities (in terms of CWE groups) referred to in commit messages and compared their numbers within the two communities.
The percentage of vulnerability mitigation commits compared to the total number of commits in projects show a growing tendency (sharper in case of JavaScript, slower for Python).
We detected 103 different CWE groups out of which 55 appeared in both languages projects.
From the eight most prevalent vulnerability types, one was mitigated by both communities in equal numbers (CWE-200), but four (CWE-20, CWE-22, CWE-79, CWE-400) was typical to JavaScript, while three (CWE-399, CWE-264, CWE-119) to Python projects.
This suggests that JavaScript and Python communities suffer the most from different types of vulnerabilities.

On the other hand, we examined the average time elapsing between the publish date of a vulnerability and the date of the commit mitigating it.
We found that in general, neither the JavaScript nor the Python community reacts very fast to appearing vulnerabilities (i.e. it takes more than 100 days on average to mitigate a vulnerability after its publish date).
However, this reaction is 1.5-14 times faster in the Python community for the most common CWE categories (even to the ones more typical to JavaScript projects), while the JavaScript community seems to take special care only of three CWE categories: CWE-200, CWE-20, and CWE-400.

\vspace{-5pt}
\begin{acks}
The presented work was carried out within the SETIT Project (2018-1.2.1-NKP-2018-00004)\footnote{Project no. 2018-1.2.1-NKP-2018-00004 has been implemented with the support provided from the National Research, Development and Innovation Fund of Hungary, financed under the 2018-1.2.1-NKP funding scheme.\vspace{-10pt}} and partially supported by grant TUDFO/47138-1/2019-ITM of the Ministry for Innovation and Technology, Hungary.
Furthermore, Péter Hegedűs was supported by the Bolyai János Scholarship of the Hungarian Academy of Sciences and the ÚNKP-19-4-SZTE-20 New National Excellence Program of the Ministry for Innovation and Technology.
\end{acks}

\bibliographystyle{ACM-Reference-Format}
\bibliography{bibl}


\begin{thebibliography}{25}


\ifx \showCODEN    \undefined \def \showCODEN     #1{\unskip}     \fi
\ifx \showDOI      \undefined \def \showDOI       #1{#1}\fi
\ifx \showISBNx    \undefined \def \showISBNx     #1{\unskip}     \fi
\ifx \showISBNxiii \undefined \def \showISBNxiii  #1{\unskip}     \fi
\ifx \showISSN     \undefined \def \showISSN      #1{\unskip}     \fi
\ifx \showLCCN     \undefined \def \showLCCN      #1{\unskip}     \fi
\ifx \shownote     \undefined \def \shownote      #1{#1}          \fi
\ifx \showarticletitle \undefined \def \showarticletitle #1{#1}   \fi
\ifx \showURL      \undefined \def \showURL       {\relax}        \fi
\providecommand\bibfield[2]{#2}
\providecommand\bibinfo[2]{#2}
\providecommand\natexlab[1]{#1}
\providecommand\showeprint[2][]{arXiv:#2}

\bibitem[\protect\citeauthoryear{Abunadi and Alenezi}{Abunadi and
  Alenezi}{2015}]%
        {10.1145/2832987.2833051}
\bibfield{author}{\bibinfo{person}{Ibrahim Abunadi} {and}
  \bibinfo{person}{Mamdouh Alenezi}.} \bibinfo{year}{2015}\natexlab{}.
\newblock \showarticletitle{Towards Cross Project Vulnerability Prediction in
  Open Source Web Applications}. In \bibinfo{booktitle}{\emph{Proceedings of
  the The International Conference on Engineering MIS 2015}} (Istanbul, Turkey)
  \emph{(\bibinfo{series}{ICEMIS ’15})}. \bibinfo{publisher}{Association for
  Computing Machinery}, \bibinfo{address}{New York, NY, USA}, Article
  \bibinfo{articleno}{42}, \bibinfo{numpages}{5}~pages.
\newblock
\showISBNx{9781450334181}
\urldef\tempurl%
\url{https://doi.org/10.1145/2832987.2833051}
\showDOI{\tempurl}


\bibitem[\protect\citeauthoryear{Anderson, Barton, B{\"o}hme, Clayton, van
  Eeten, Levi, Moore, and Savage}{Anderson et~al\mbox{.}}{2013}]%
        {Anderson2013}
\bibfield{author}{\bibinfo{person}{Ross Anderson}, \bibinfo{person}{Chris
  Barton}, \bibinfo{person}{Rainer B{\"o}hme}, \bibinfo{person}{Richard
  Clayton}, \bibinfo{person}{Michel J.~G. van Eeten}, \bibinfo{person}{Michael
  Levi}, \bibinfo{person}{Tyler Moore}, {and} \bibinfo{person}{Stefan Savage}.}
  \bibinfo{year}{2013}\natexlab{}.
\newblock \bibinfo{booktitle}{\emph{Measuring the Cost of Cybercrime}}.
\newblock \bibinfo{publisher}{Springer Berlin Heidelberg},
  \bibinfo{address}{Berlin, Heidelberg}, \bibinfo{pages}{265--300}.
\newblock
\showISBNx{978-3-642-39498-0}
\urldef\tempurl%
\url{https://doi.org/10.1007/978-3-642-39498-0_12}
\showDOI{\tempurl}


\bibitem[\protect\citeauthoryear{CVE}{CVE}{2020}]%
        {cve}
CVE \bibinfo{year}{2020}\natexlab{}.
\newblock \bibinfo{booktitle}{\emph{{Common Vulnerabilities and Exposures}}}.
\newblock
\urldef\tempurl%
\url{https://cve.mitre.org/}
\showURL{%
\tempurl}


\bibitem[\protect\citeauthoryear{CVE Manager}{CVE Manager}{2020}]%
        {cve-manager}
CVE Manager \bibinfo{year}{2020}\natexlab{}.
\newblock \bibinfo{booktitle}{\emph{{CVE Manager}}}.
\newblock
\urldef\tempurl%
\url{https://github.com/aatlasis/cve_manager}
\showURL{%
\tempurl}


\bibitem[\protect\citeauthoryear{CWE}{CWE}{2020}]%
        {cwe}
CWE \bibinfo{year}{2020}\natexlab{}.
\newblock \bibinfo{booktitle}{\emph{{Common Weaknesses Enumeration}}}.
\newblock
\urldef\tempurl%
\url{https://cwe.mitre.org/}
\showURL{%
\tempurl}


\bibitem[\protect\citeauthoryear{Frei, May, Fiedler, and Plattner}{Frei
  et~al\mbox{.}}{2006}]%
        {10.1145/1162666.1162671}
\bibfield{author}{\bibinfo{person}{Stefan Frei}, \bibinfo{person}{Martin May},
  \bibinfo{person}{Ulrich Fiedler}, {and} \bibinfo{person}{Bernhard Plattner}.}
  \bibinfo{year}{2006}\natexlab{}.
\newblock \showarticletitle{Large-Scale Vulnerability Analysis}. In
  \bibinfo{booktitle}{\emph{Proceedings of the 2006 SIGCOMM Workshop on
  Large-Scale Attack Defense}} (Pisa, Italy) \emph{(\bibinfo{series}{LSAD
  ’06})}. \bibinfo{publisher}{Association for Computing Machinery},
  \bibinfo{address}{New York, NY, USA}, \bibinfo{pages}{131–138}.
\newblock
\showISBNx{1595935711}
\urldef\tempurl%
\url{https://doi.org/10.1145/1162666.1162671}
\showDOI{\tempurl}


\bibitem[\protect\citeauthoryear{Gkortzis, Mitropoulos, and Spinellis}{Gkortzis
  et~al\mbox{.}}{2018}]%
        {10.1145/3196398.3196454}
\bibfield{author}{\bibinfo{person}{Antonios Gkortzis},
  \bibinfo{person}{Dimitris Mitropoulos}, {and} \bibinfo{person}{Diomidis
  Spinellis}.} \bibinfo{year}{2018}\natexlab{}.
\newblock \showarticletitle{VulinOSS: A Dataset of Security Vulnerabilities in
  Open-Source Systems}. In \bibinfo{booktitle}{\emph{Proceedings of the 15th
  International Conference on Mining Software Repositories}} (Gothenburg,
  Sweden) \emph{(\bibinfo{series}{MSR ’18})}. \bibinfo{publisher}{Association
  for Computing Machinery}, \bibinfo{address}{New York, NY, USA},
  \bibinfo{pages}{18–21}.
\newblock
\showISBNx{9781450357166}
\urldef\tempurl%
\url{https://doi.org/10.1145/3196398.3196454}
\showDOI{\tempurl}


\bibitem[\protect\citeauthoryear{Li and Paxson}{Li and Paxson}{2017}]%
        {frankli2017}
\bibfield{author}{\bibinfo{person}{Frank Li} {and} \bibinfo{person}{Vern
  Paxson}.} \bibinfo{year}{2017}\natexlab{}.
\newblock \showarticletitle{A Large-Scale Empirical Study of Security Patches}.
  In \bibinfo{booktitle}{\emph{Proceedings of the 2017 ACM SIGSAC Conference on
  Computer and Communications Security}} (Dallas, Texas, USA)
  \emph{(\bibinfo{series}{CCS ’17})}. \bibinfo{publisher}{Association for
  Computing Machinery}, \bibinfo{address}{New York, NY, USA},
  \bibinfo{pages}{2201–2215}.
\newblock
\showISBNx{9781450349468}
\urldef\tempurl%
\url{https://doi.org/10.1145/3133956.3134072}
\showDOI{\tempurl}


\bibitem[\protect\citeauthoryear{Massacci and Nguyen}{Massacci and
  Nguyen}{2010}]%
        {10.1145/1853919.1853925}
\bibfield{author}{\bibinfo{person}{Fabio Massacci} {and}
  \bibinfo{person}{Viet~Hung Nguyen}.} \bibinfo{year}{2010}\natexlab{}.
\newblock \showarticletitle{Which is the Right Source for Vulnerability
  Studies? An Empirical Analysis on Mozilla Firefox}. In
  \bibinfo{booktitle}{\emph{Proceedings of the 6th International Workshop on
  Security Measurements and Metrics}} (Bolzano, Italy)
  \emph{(\bibinfo{series}{MetriSec ’10})}. \bibinfo{publisher}{Association
  for Computing Machinery}, \bibinfo{address}{New York, NY, USA}, Article
  \bibinfo{articleno}{4}, \bibinfo{numpages}{8}~pages.
\newblock
\showISBNx{9781450303408}
\urldef\tempurl%
\url{https://doi.org/10.1145/1853919.1853925}
\showDOI{\tempurl}


\bibitem[\protect\citeauthoryear{McKinney et~al\mbox{.}}{McKinney
  et~al\mbox{.}}{2011}]%
        {mckinney2011pandas}
\bibfield{author}{\bibinfo{person}{Wes McKinney} {et~al\mbox{.}}}
  \bibinfo{year}{2011}\natexlab{}.
\newblock \showarticletitle{{Pandas: a Foundational Python Library for Data
  Analysis and Statistics}}.
\newblock \bibinfo{journal}{\emph{Python for High Performance and Scientific
  Computing}} \bibinfo{volume}{14}, \bibinfo{number}{9} (\bibinfo{year}{2011}).
\newblock


\bibitem[\protect\citeauthoryear{Node.js}{Node.js}{2020}]%
        {nodejsdocumentation}
Node.js \bibinfo{year}{2020}\natexlab{}.
\newblock \bibinfo{booktitle}{\emph{{Modules | Node.js v13.7.0
  Documentation}}}.
\newblock
\urldef\tempurl%
\url{https://nodejs.org/api/modules.html#modules_folders_as_modules}
\showURL{%
\tempurl}


\bibitem[\protect\citeauthoryear{NVD}{NVD}{2020}]%
        {nvd}
NVD \bibinfo{year}{2020}\natexlab{}.
\newblock \bibinfo{booktitle}{\emph{{National Vulnerability Database}}}.
\newblock
\urldef\tempurl%
\url{https://nvd.nist.gov/}
\showURL{%
\tempurl}


\bibitem[\protect\citeauthoryear{P{\'e}ter, B{\'e}la, Andrea, Davood,
  {\'A}rp{\'a}d, Rudolf, and Ali}{P{\'e}ter et~al\mbox{.}}{2019}]%
        {GVS19}
\bibfield{author}{\bibinfo{person}{Gyimesi P{\'e}ter},
  \bibinfo{person}{Vancsics B{\'e}la}, \bibinfo{person}{Stocco Andrea},
  \bibinfo{person}{Mazinanian Davood}, \bibinfo{person}{Besz{\'e}des
  {\'A}rp{\'a}d}, \bibinfo{person}{Ferenc Rudolf}, {and}
  \bibinfo{person}{Mesbah Ali}.} \bibinfo{year}{2019}\natexlab{}.
\newblock \showarticletitle{BugsJS: A Benchmark of JavaScript Bugs}. In
  \bibinfo{booktitle}{\emph{Proceedings of the 12th IEEE Conference on Software
  Testing, Validation and Verification (ICST)}}. IEEE,
  \bibinfo{pages}{90--101}.
\newblock
\urldef\tempurl%
\url{https://doi.org/10.1109/ICST.2019.00019}
\showDOI{\tempurl}


\bibitem[\protect\citeauthoryear{Pietri, Spinellis, and Zacchiroli}{Pietri
  et~al\mbox{.}}{2019}]%
        {MSR19SH}
\bibfield{author}{\bibinfo{person}{Antoine Pietri}, \bibinfo{person}{Diomidis
  Spinellis}, {and} \bibinfo{person}{Stefano Zacchiroli}.}
  \bibinfo{year}{2019}\natexlab{}.
\newblock \showarticletitle{The Software Heritage Graph Dataset: Public
  software development under one roof}. In \bibinfo{booktitle}{\emph{MSR 2019:
  The 16th International Conference on Mining Software Repositories}}.
  \bibinfo{publisher}{IEEE}, \bibinfo{pages}{138--142}.
\newblock
\urldef\tempurl%
\url{https://doi.org/10.1109/MSR.2019.00030}
\showDOI{\tempurl}


\bibitem[\protect\citeauthoryear{Pietri, Spinellis, and Zacchiroli}{Pietri
  et~al\mbox{.}}{2020}]%
        {MSR20DC}
\bibfield{author}{\bibinfo{person}{Antoine Pietri}, \bibinfo{person}{Diomidis
  Spinellis}, {and} \bibinfo{person}{Stefano Zacchiroli}.}
  \bibinfo{year}{2020}\natexlab{}.
\newblock \showarticletitle{The {Software Heritage Graph Dataset}: Large-scale
  Analysis of Public Software Development History}. In
  \bibinfo{booktitle}{\emph{MSR 2020: The 17th International Conference on
  Mining Software Repositories}}. \bibinfo{publisher}{IEEE}.
\newblock


\bibitem[\protect\citeauthoryear{Pilgrim and Willison}{Pilgrim and
  Willison}{2009}]%
        {pilgrim2009dive}
\bibfield{author}{\bibinfo{person}{Mark Pilgrim} {and} \bibinfo{person}{Simon
  Willison}.} \bibinfo{year}{2009}\natexlab{}.
\newblock \bibinfo{booktitle}{\emph{Dive Into Python 3}}.
  Vol.~\bibinfo{volume}{2}.
\newblock \bibinfo{publisher}{Springer}.
\newblock


\bibitem[\protect\citeauthoryear{PyPI}{PyPI}{2020}]%
        {pypi}
PyPI \bibinfo{year}{2020}\natexlab{}.
\newblock \bibinfo{booktitle}{\emph{{Python Package Index}}}.
\newblock
\urldef\tempurl%
\url{https://pypi.org/}
\showURL{%
\tempurl}


\bibitem[\protect\citeauthoryear{Python Packaging User Guide}{Python Packaging
  User Guide}{2020}]%
        {packagingpython}
Python Packaging User Guide \bibinfo{year}{2020}\natexlab{}.
\newblock \bibinfo{booktitle}{\emph{{Packaging and distributing projects -
  Python Packaging User Guide}}}.
\newblock
\urldef\tempurl%
\url{https://packaging.python.org/tutorials/packaging-projects}
\showURL{%
\tempurl}


\bibitem[\protect\citeauthoryear{{Shahzad}, {Shafiq}, and {Liu}}{{Shahzad}
  et~al\mbox{.}}{2012}]%
        {6227141}
\bibfield{author}{\bibinfo{person}{M. {Shahzad}}, \bibinfo{person}{M.~Z.
  {Shafiq}}, {and} \bibinfo{person}{A.~X. {Liu}}.}
  \bibinfo{year}{2012}\natexlab{}.
\newblock \showarticletitle{{A Large Scale Exploratory Analysis of Software
  Vulnerability Life Cycles}}. In \bibinfo{booktitle}{\emph{Proceedings of the
  34th International Conference on Software Engineering (ICSE)}}.
  \bibinfo{pages}{771--781}.
\newblock
\showISSN{0270-5257}
\urldef\tempurl%
\url{https://doi.org/10.1109/ICSE.2012.6227141}
\showDOI{\tempurl}


\bibitem[\protect\citeauthoryear{\'Sliwerski, Zimmermann, and
  Zeller}{\'Sliwerski et~al\mbox{.}}{2005}]%
        {10.1145/1083142.1083147}
\bibfield{author}{\bibinfo{person}{Jacek \'Sliwerski}, \bibinfo{person}{Thomas
  Zimmermann}, {and} \bibinfo{person}{Andreas Zeller}.}
  \bibinfo{year}{2005}\natexlab{}.
\newblock \showarticletitle{When Do Changes Induce Fixes?}. In
  \bibinfo{booktitle}{\emph{Proceedings of the 2005 International Workshop on
  Mining Software Repositories}} (St. Louis, Missouri)
  \emph{(\bibinfo{series}{MSR ’05})}. \bibinfo{publisher}{Association for
  Computing Machinery}, \bibinfo{address}{New York, NY, USA},
  \bibinfo{pages}{1–5}.
\newblock
\showISBNx{1595931236}
\urldef\tempurl%
\url{https://doi.org/10.1145/1083142.1083147}
\showDOI{\tempurl}


\bibitem[\protect\citeauthoryear{Sudo Vulnerability}{Sudo
  Vulnerability}{2020}]%
        {sudo-macos}
Sudo Vulnerability \bibinfo{year}{2020}\natexlab{}.
\newblock \bibinfo{booktitle}{\emph{{Sudo vulnerability in macOS}}}.
\newblock
\urldef\tempurl%
\url{https://www.techradar.com/news/linux-and-macos-pcs-hit-by-serious-sudo-vulnerability}
\showURL{%
\tempurl}


\bibitem[\protect\citeauthoryear{Symfony Docs}{Symfony Docs}{2020}]%
        {symfonydocs}
Symfony Docs \bibinfo{year}{2020}\natexlab{}.
\newblock \bibinfo{booktitle}{\emph{{Installing Encore (Symfony Docs)}}}.
\newblock
\urldef\tempurl%
\url{https://symfony.com/doc/current/frontend/encore/installation.html#installing-encore-in-non-symfony-applications}
\showURL{%
\tempurl}


\bibitem[\protect\citeauthoryear{V{\'a}squez, Bavota, and
  Escobar-Velasquez}{V{\'a}squez et~al\mbox{.}}{2017}]%
        {Vsquez2017AnES}
\bibfield{author}{\bibinfo{person}{Mario~Linares V{\'a}squez},
  \bibinfo{person}{Gabriele Bavota}, {and} \bibinfo{person}{Camilo
  Escobar-Velasquez}.} \bibinfo{year}{2017}\natexlab{}.
\newblock \showarticletitle{An Empirical Study on Android-Related
  Vulnerabilities}.
\newblock \bibinfo{journal}{\emph{Proceedings of the IEEE/ACM 14th
  International Conference on Mining Software Repositories (MSR)}}
  (\bibinfo{year}{2017}), \bibinfo{pages}{2--13}.
\newblock


\bibitem[\protect\citeauthoryear{{Xu}, {Chen}, {Chandramohan}, {Liu}, and
  {Song}}{{Xu} et~al\mbox{.}}{2017}]%
        {xu2017spain}
\bibfield{author}{\bibinfo{person}{Z. {Xu}}, \bibinfo{person}{B. {Chen}},
  \bibinfo{person}{M. {Chandramohan}}, \bibinfo{person}{Y. {Liu}}, {and}
  \bibinfo{person}{F. {Song}}.} \bibinfo{year}{2017}\natexlab{}.
\newblock \showarticletitle{SPAIN: Security Patch Analysis for Binaries towards
  Understanding the Pain and Pills}. In \bibinfo{booktitle}{\emph{Proceedings
  of the IEEE/ACM 39th International Conference on Software Engineering
  (ICSE)}}. \bibinfo{pages}{462--472}.
\newblock
\showISSN{1558-1225}
\urldef\tempurl%
\url{https://doi.org/10.1109/ICSE.2017.49}
\showDOI{\tempurl}


\bibitem[\protect\citeauthoryear{Younker}{Younker}{2009}]%
        {younker2009foundations}
\bibfield{author}{\bibinfo{person}{Jeff Younker}.}
  \bibinfo{year}{2009}\natexlab{}.
\newblock \bibinfo{booktitle}{\emph{Foundations of agile python development}}.
\newblock \bibinfo{publisher}{Apress}.
\newblock


\end{thebibliography}

\end{document}